\documentstyle[graphicx,epsfig]{aipproc}

\newcommand{\E}[1]{$10^{#1}$eV}

\newcommand{\AmS}{{\protect\the\textfont2
  A\kern-.1667em\lower.5ex\hbox{M}\kern-.125emS}}

\begin{document}
\title{Establishing The GZK Cutoff With Ultra High Energy Tau Neutrinos}
\author{A. Letessier-Selvon\thanks{Presented at the International Workshop on Observing Ultra High Energy Cosmic 
Rays From Space and Earth. August 9-12, 2000, Puebla (Mexico).}}
\address{Laboratoire de Physique Nucl\'eaire et des Hautes \'Energies,\\
IN2P3-CNRS, Universit\'es Paris 6 \& 7,\\
4 place Jussieu, Tour 33 RdC, 75252 Paris Cedex 05 France\\
e-mail : Antoine.Letessier-Selvon@in2p3.fr}%

\maketitle

\begin{abstract}
The cosmic ray spectrum has been shown to extend well beyond \E{20}. With nearly 20 events observed 
in the last 40 years, it is now established that particles are accelerated or produced in the 
universe with energy near \E{21}. In all production models neutrinos and photons are  
part of the cosmic ray flux. In acceleration models (bottom-up models), they are produced as 
secondaries of the possible interactions of the accelerated charged particle, in direct production models 
(top-down models)  they are a dominant fraction of the decay chain. 
In addition, hadrons above the GZK threshold energy will also produce, 
along their path in the Universe, neutrinos and photons as secondaries of the pion photo-production
processes. Therefore, photons and in particular neutrinos, are very distinctive signatures of the 
nature and distribution of the potential sources of ultra high energy cosmic rays. 
In the following we expose the identification capabilities of the Auger observatories. In the hypothesis of 
\mbox{$\nu_\mu\rightarrow\nu_\tau$} oscillations with full mixing, special emphasis
is placed on the observation of tau neutrinos, with which Auger is able to establish the GZK cutoff 
as well as to provide a strong and  model independant constraint on the top-down 
sources of ultra high energy cosmic rays.
\vspace{1pc}
\end{abstract}
\maketitle

\vspace*{-1cm}
\section{INTRODUCTION}
The cosmic ray spectrum is now proved\cite{HiresTaup99,Agasa00} to extend beyond 
\E{20}. To be observed on Earth with such energies, particles must be produced or accelerated
in the Universe  with energy near or above \E{21}. Conventional acceleration mechanisms in
astrophysical objects can only reach this requirement by stretching to the limit their available 
parameter space, making such scenarios unlikely to explain the origin of UHECR. 
Alternative hypotheses involving collapse of Topological Defects (TD) or decay of Super Massive Relic Particles (SMRP)
are well suited to produce particles above \E{20} but need a proof of existence. 
\par
Transport, from the source to Earth, is also an issue. At those extreme energies the  Cosmic Microwave 
Background Radiation makes the Universe essentially opaque to protons, nuclei and photons which
suffer energy losses from pion photo production, photo-disintegration or pair production. 
These processes led Greisen, Zatsepin and Kuzmin\cite{GZK} to predict a spectral cutoff in the cosmic ray 
spectrum around $5\times$\E{19}, the GZK cutoff. The available data although still very scarce do 
not support the existence of such a cutoff. Therefore the sources are either close by and locally
over dense for the cutoff not to show, or some new physics prevent the UHECR from the expected energy 
losses against the CMB photons.
\par
The following will briefly develop the arguments mentioned in this introduction. Interested readers should consult
the numerous reviews devoted to the subject\cite{Yoshida,Sigl,Bertou} for more details.

\section{OBSERVATIONS}
\begin{figure}[!htb]
\begin{center}
\includegraphics*[width=10cm]{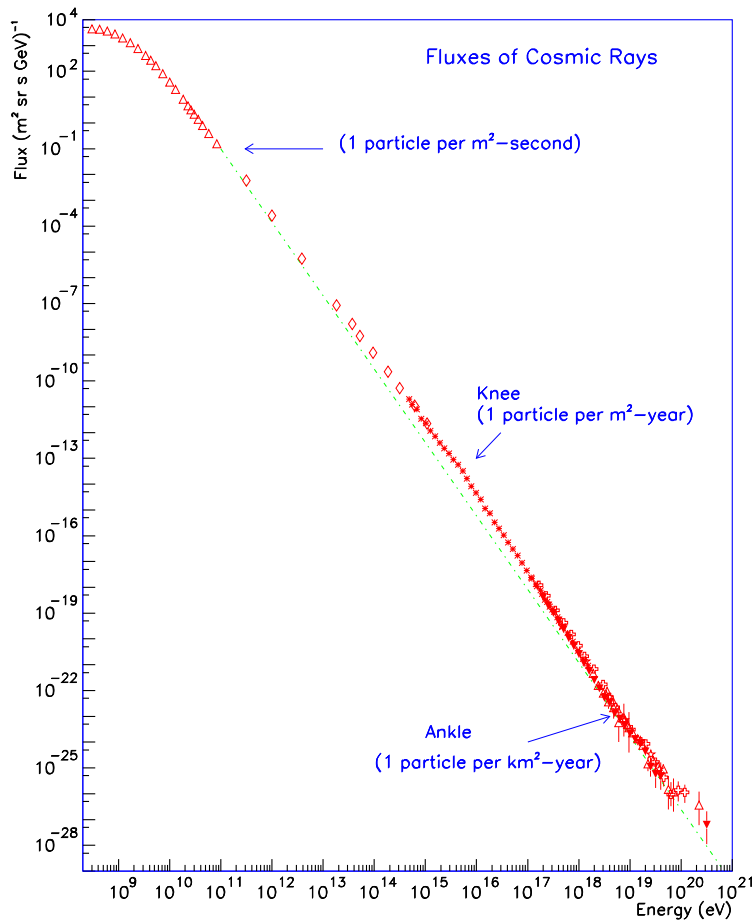}
\caption{The cosmic rays spectrum}
\label{spectrum}
\end{center}
\end{figure}

The differential spectrum of cosmic ray flux\cite{Swordy} as a function of energy is shown on Figure~\ref{spectrum}. 
Integrated fluxes above three energy values are indicated: 1 particle/m$^2$-second
above 1~TeV, 1 particle/m$^2$-year above 10~PeV, 1 particle/km$^2$-year above 10
EeV.  The energy spectrum is surprisingly regular
in shape. From the GeV energies to the GZK cutoff, it can be represented
simply by three power-law segments interrupted by two breaks, the so-called
``knee" and ``ankle". 
\par
\begin{figure}[!htb]
\begin{center}
\includegraphics*[width=10cm]{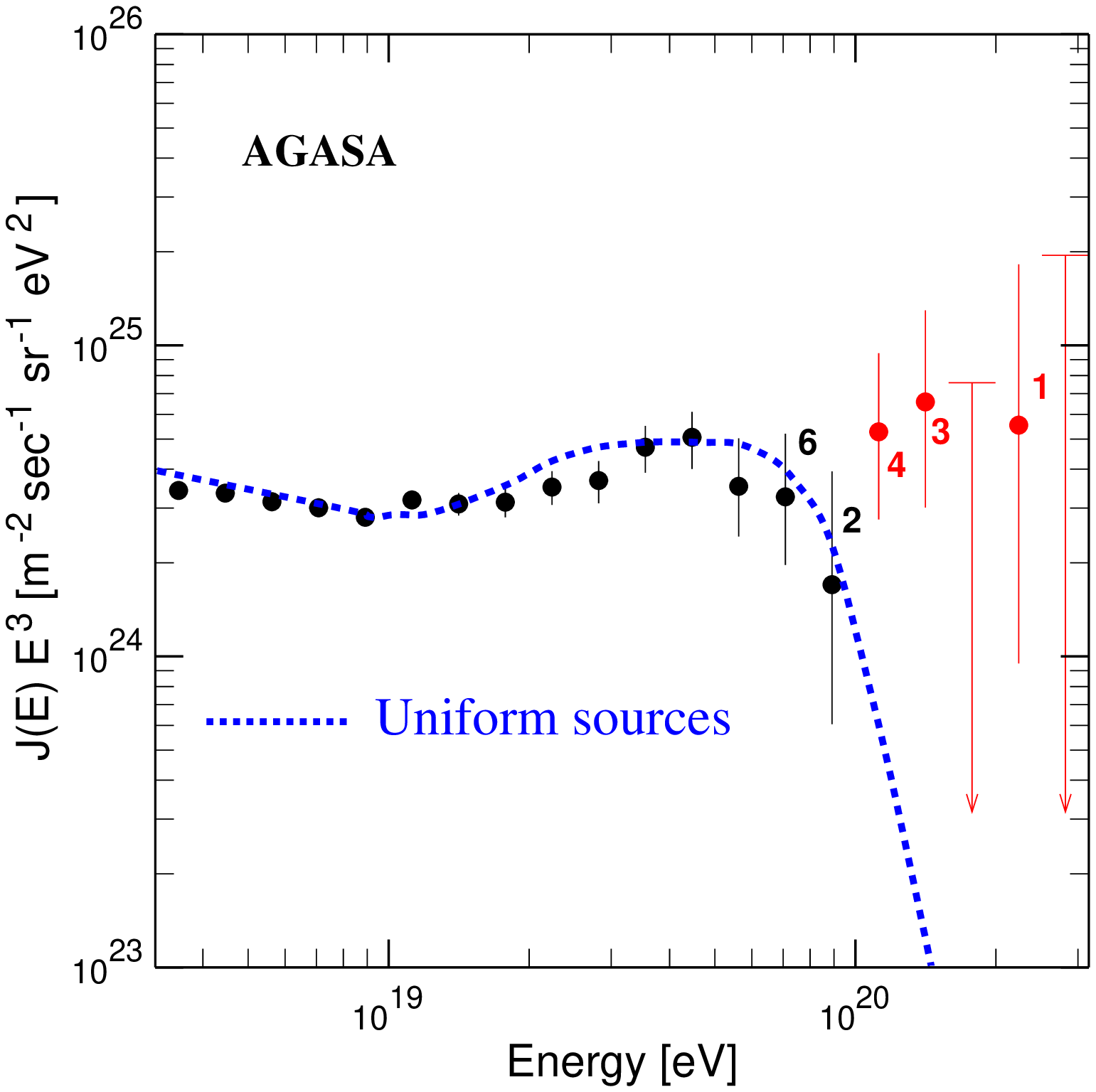}
\caption{Highest energy region of the cosmic ray spectrum as observed by the
AGASA detector.
 The figures near the data points indicate the
number of events in the corresponding energy bin. The arrows show 90\%
confidence level upper limits. The dashed line is the expected spectrum if the
sources were cosmologically distributed.}\label{takeda}
\end{center}
\end{figure}

Figure~\ref{takeda} is a zoom on the
highest energy part of the total spectrum where only the
latest AGASA data\cite{Agasa00} is  displayed.
 On this figure, the energy spectrum is multiplied by $E^3$ so
that the part below the EeV energies becomes flat. 
Comparing the data points and the dashed line one 
 has a clear view of what can be expected from a cosmological (uniform) distribution of conventional
 sources and what is observed.
 
The cutoff, that would be expected if the sources were cosmologically and uniformly
distributed and if the observed cosmic rays had no exotic propagation or interaction
properties, \emph{is not} present in the observed data.
 
\par
In the search for potential sources, one
looks for correlations of the UHECR arrival directions with the distribution of matter within a
few tens of Mpc. 
Such an analysis 
was done by the AGASA experiment for the highest energy range~\cite{Takeda2}.
No convincing deviation from isotropy was found. 
\par
If the sources of UHECR are nearby astrophysical objects and if, as
expected, they are in small numbers,
a selection of the events with the largest magnetic rigidity would combine
into multiplets (cluster of events whose error boxes overlap).

\begin{figure}[t] 
\begin{center}  
\epsfig{file=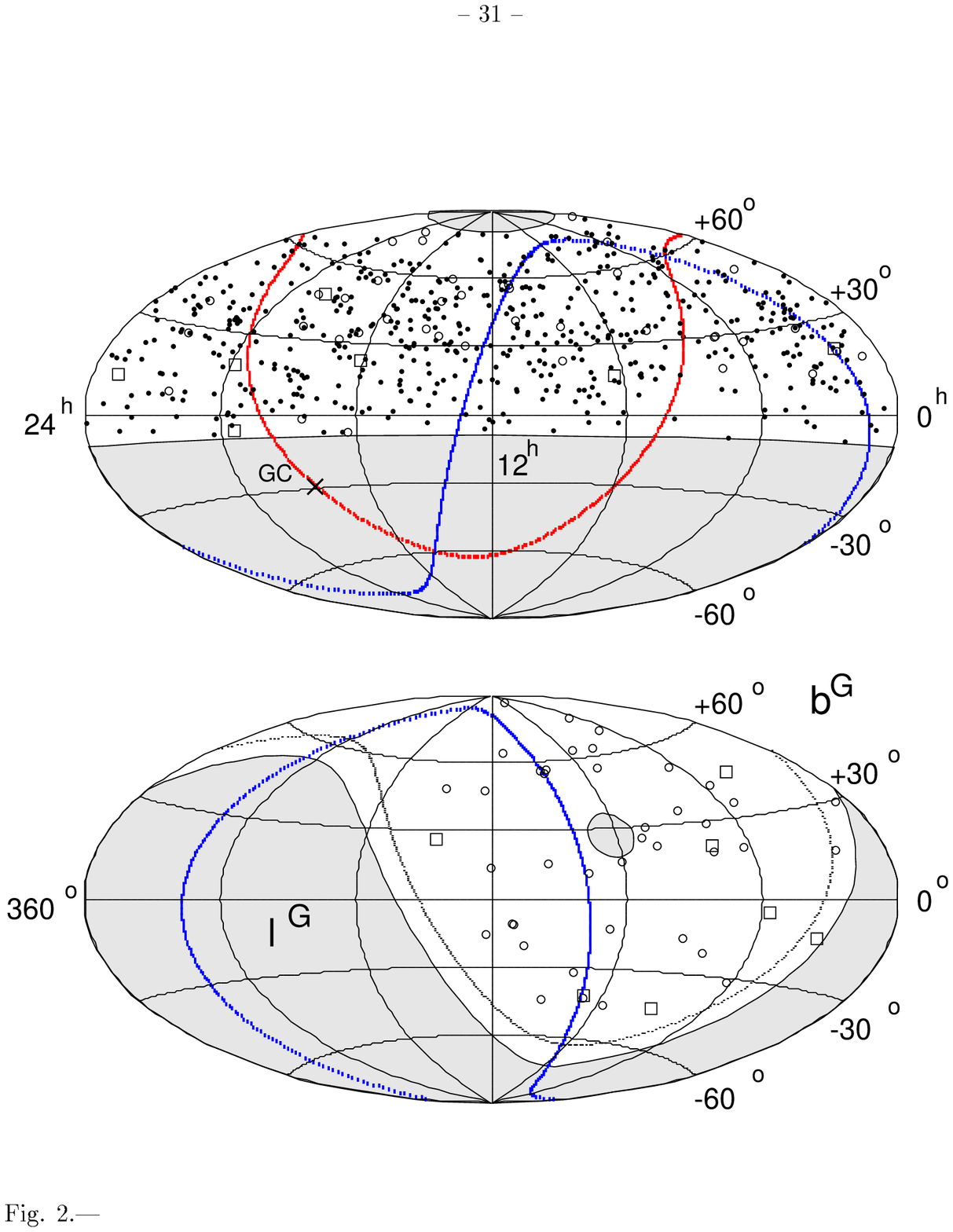,bbllx=71pt,bblly=187pt,bburx=528pt,%
bbury=398pt,width=10cm,clip=} 
\caption{Arrival directions (galactic coordinates) of cosmic rays with $E>$40~EeV,
AGASA\label{bigevents2}.}
\end{center}   
\end{figure}

Figure \ref{bigevents2} shows the subsample of events in the AGASA catalog with
energies in excess of 100~EeV (squares) and in the range 40-100~EeV (circles).
One can see that there are three doublets and one triplet. 
The chance probability of having as many multiplets as observed with
a uniform distribution is estimated to be less than 1\%~
\par
The non uniform sky coverage -all present detectors are in the northern hemisphere- and the small statistics available
make anisotropy studies difficult.
The Auger observatories are designed with full sky coverage and large detection areas to overcome these  
difficulties.

\section{TRANSPORT AND PRODUCTION}
Today's understanding of the phenomena responsible for the production of UHECR 
is still limited. One distinguishes two classes of processes: the ``Top-Down''
and ``Bottom-Up'' scenarios. In Top-Down scenarios, the cosmic ray is a decay products of a 
super-massive particle. Such particles  with masses exceeding \E{21} are either meta-stable relics 
of some primordial field or GUT gauge bosons produced by the radiation or collapse of topological defects.  
In the Bottom-Up scenarios, the energy is transferred 
to a charged particle at rest through its electromagnetic interactions.
This classical approach does not require new physics.

\par
At energies above 10~EeV and except for neutrinos, the Universe 
is not transparent to ordinary stable particles on scales larger than about 100 Mpc. 
Regardless of their nature, cosmic rays lose energy in their interaction with the various photon 
backgrounds, dominantly the copious Cosmic Microwave Background (CMB) but also 
Infra-Red and Radio. 
The absence of prominent visible astrophysical objects in the direction of the 
observed highest  energy cosmic rays together with this distance limitation adds severe
constraints on the ``classical'' Bottom-Up picture.

\subsection{The GZK cutoff}
The energy at which the Greisen-Zatsepin-Kuzmin (GZK) cutoff takes place is given by
the threshold for pion photo-production in the collisions of protons and CMB-photons.
For an average CMB photon (\E{-3}), one obtains $E_{th}=7 \times $\E{19}.
The interaction length can be estimated from the pion photo-production 
cross section 
and the CMB density~:
\mbox{$ L=(\sigma \rho)^{-1}\simeq 6\,\mbox{Mpc}$}.       
\par
\begin{figure}[!t]
\begin{minipage}[b]{0.47\linewidth}  
\centering \includegraphics[width=\linewidth]{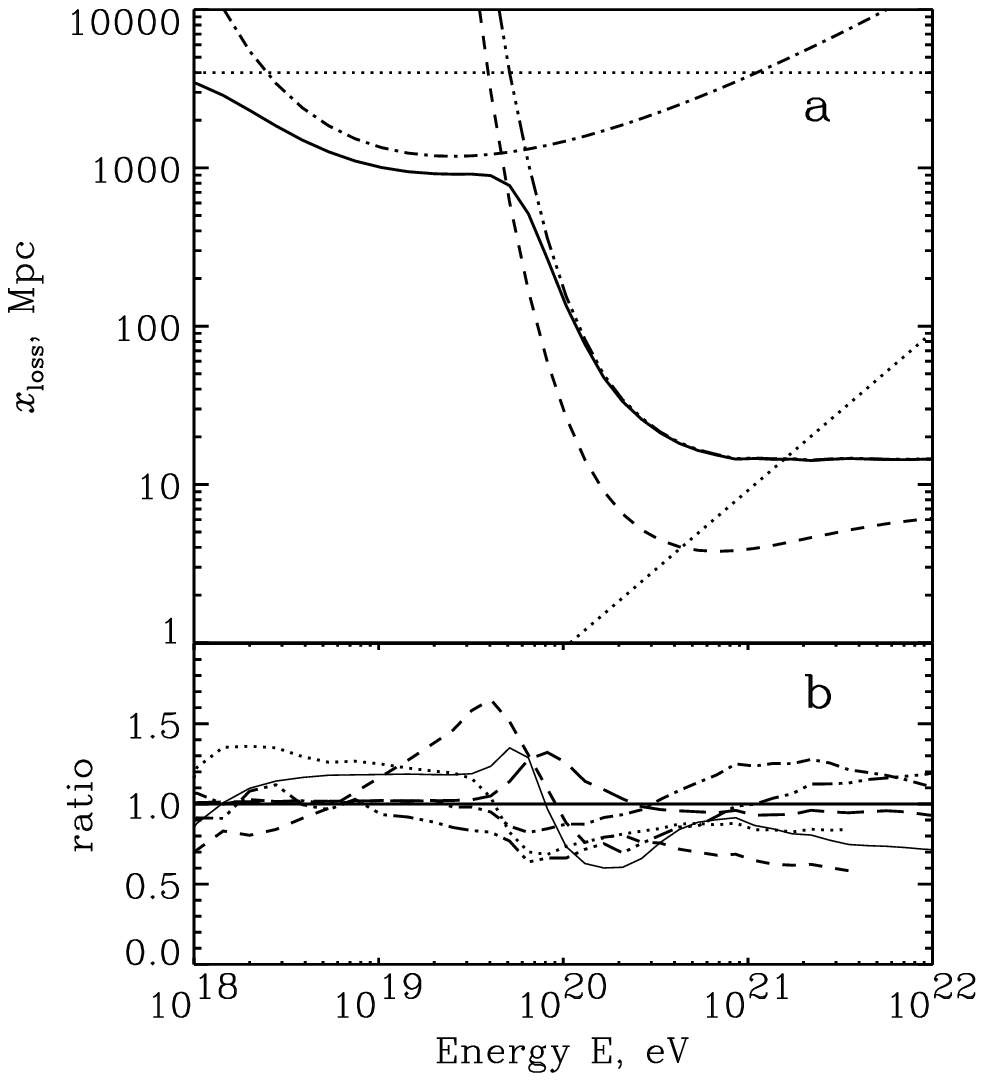}
\caption{a) Loss length of a proton from a recent complete monte calo simulation by  T.~Stanev et al.~(astro-ph/0003484).\newline
b) Ratios with other calculations.}\label{GZK-loss}
\end{minipage}
\begin{minipage}[b]{0.47\linewidth}
\centering \includegraphics[width=\linewidth]{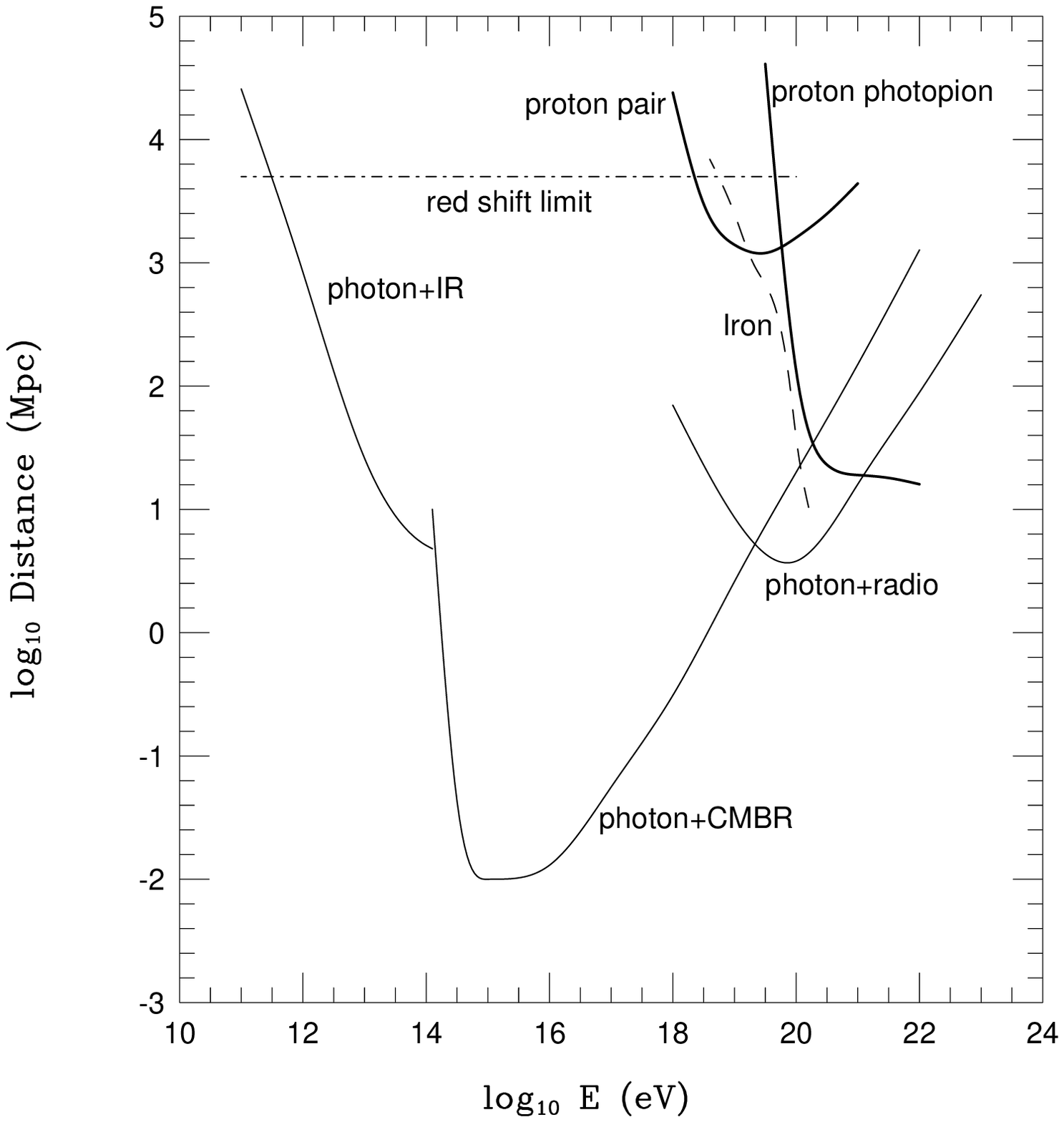}
\caption{Attenuation length of photons, protons and iron
as a function of energy. \label{adrf24}  
Double pair production (not shown) limits the photon attenuation length to about 100 Mpc above \E{22}.}
\end{minipage}
\end{figure}

A recent Monte Carlo~\cite{Stanev00} calculation, including red shift, pair production and pion 
photo-production losses, is shown on Figure~\ref{GZK-loss}. The loss length $x_{loss}$
is defined as  $ x_{loss}=\frac{E}{dE/dx}$. 
Above 100 EeV photo-production processes are dominant and the loss length falls below 13 Mpc.

\par
For nuclei, the situation is usually worse. They photo-disintegrate in the 
CMB and infrared radiations losing on average 3 to 4 nucleons per Mpc when their energy 
exceeds 2$\times$\E{19} to 2$\times$\E{20} depending on the IR background density value. 
\par
Top-Down production mechanisms predict that, at the source, photons and neutrinos
dominate over ordinary hadrons by about a factor four to ten\cite{Sigl,Sarkar}. 
An observed dominance of gammas or neutrinos in the supra-GZK range would then be an  
inescapable signature of a super-heavy particle decay or TD interaction. 
High energy photons traveling through the Universe produce $e^+e^-$ pairs when colliding with the 
Infra-Red/Optical (IR/O), CMB or Universal Radio Background (URB) photons. As can be seen on 
Figure~\ref{adrf24} the attenuation length gets below 10 Mpc for photon energies between $3\times$\E{13} 
and \E{20}. In this energy range
the Universe is opaque to photons on cosmological scales. 

Once the photon has converted, the $e^+e^-$ pair will in turn produce photons mostly via Inverse Compton 
Scattering (ICS). Those two dominant processes  are responsible for the production of 
electromagnetic (EM) cascades.
On Figure~\ref{adrf24} one sees that, at the pair production threshold on the CMB photons 
($2\times$\E{14}), conversion occurs on distances of about 10~kpc (a thousand times smaller than for protons at 
GZK energies) while subsequent ICS of electrons on the 
CMB in the Thomson regime will occur on even smaller scales (1~kpc). 
\par
As a consequence, most photons of ultra high energy  will produce, through successive collisions
on the various photon backgrounds (URB, CMB, IR/O), lower and lower energy cascades 
and pile up in the form of a diffuse 
photon background below \E{12} with a typical  power law spectrum of index $\alpha=1.5$.  
This is a very important fact as measurements of the diffuse gamma ray background in the $10^7$-\E{11} 
range done for example by EGRET\cite{EGRET} will impose limits on the photon production fluxes of 
Top-Down mechanisms and consequently on the abundance of  topological defects or relic super-heavy 
particles.

Neutrinos are the only known particle that can travel through space unaffected even on large distances,
carrying intact the properties of the source to the observer.
They may prove to be an unambiguous signature of the new physics underlying the 
production mechanisms.
\subsection{Bottom-Up acceleration}
\begin{figure}[!t]
\begin{center}
\includegraphics[width=10cm]{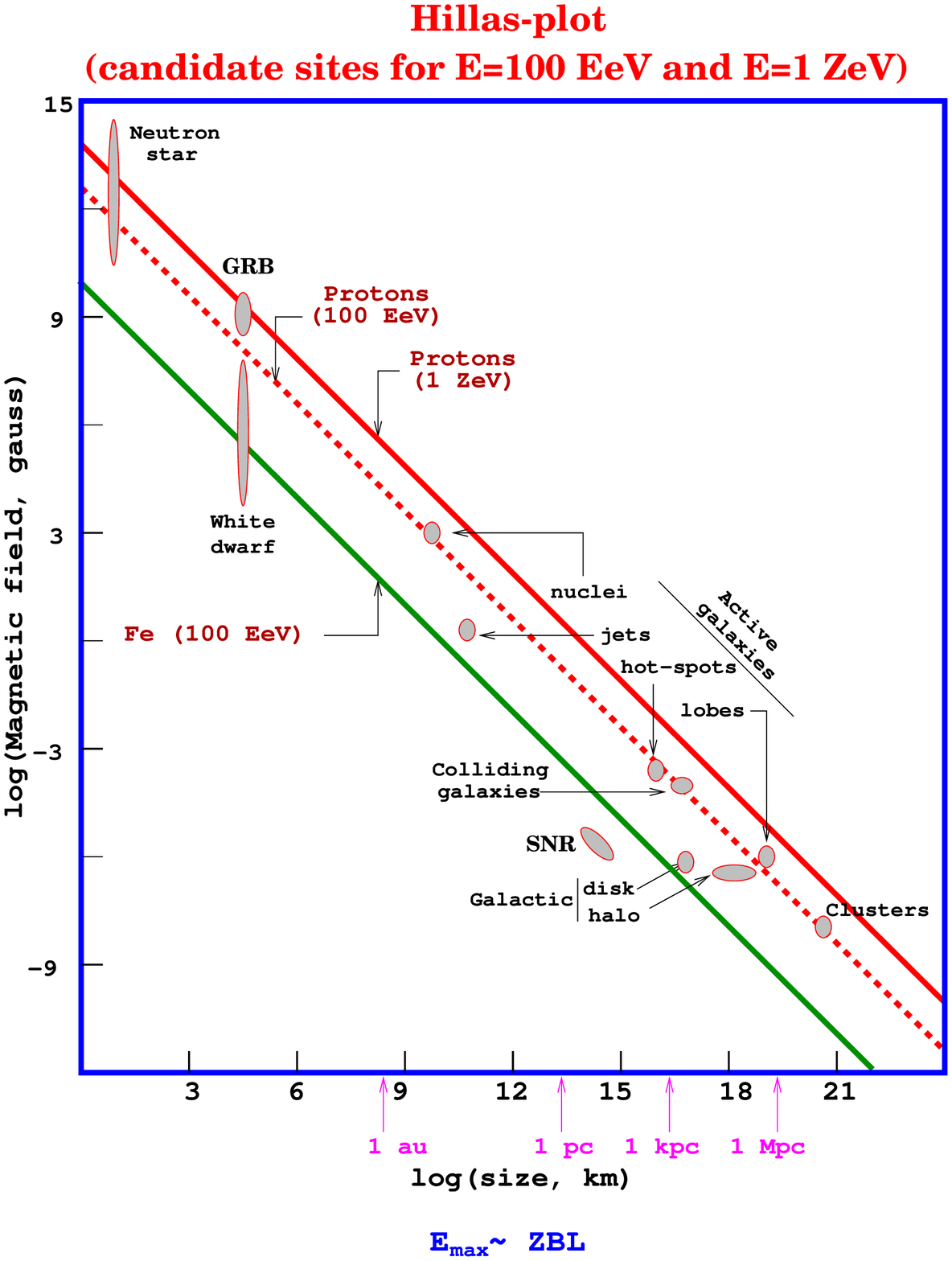}
\caption{Size and magnetic field strength of possible acceleration sites. Objects below the diagonal 
lines cannot accelerate the corresponding elements above \E{20} or \E{21}.\label{Hillas-Diagram}}
\end{center}
\end{figure}

In the conventional acceleration scenarios one distinguishes two types
of mechanisms~:
\begin{itemize}
\item Direct acceleration by very high electric fields in or near 
very compact objects. This does not naturally provide a power-law spectrum. 
\item Diffusive shock acceleration in all systems where shock waves are present.
This statistical acceleration, known as the Fermi mechanism, 
naturally provides a power law spectrum.
\end{itemize} 
Hillas has shown\cite{Hillas} that irrespective of the details of the acceleration 
mechanisms, the maximum energy of a particle of charge $Ze$ within a given site of size $R$ is: 
\begin{equation}
E_{\text{max}}\approx\beta Z\left(\frac{B}{1\,\mu \text{G}}\right )\left(\frac{R}{1\,\text{kpc}}\right)10^{18}~\text{eV}
\label{eq:Hillas}
\end{equation}
where $B$ is the magnetic field inside the acceleration volume and $\beta$ the velocity of the shock wave 
or the efficiency of the acceleration mechanism. This condition 
is nicely represented by the Hillas diagram shown in Figure \ref{Hillas-Diagram}.
Inspecting this diagram one sees that only a few astrophysical sources satisfy the necessary
condition given by Eq.~(\ref{eq:Hillas}). 

Let us briefly review them~:
\par{\it \bf Pulsars}~:
From a dimensional analysis, one obtains up to $10^{20}$ volts for the potential drop in a rotating magnetic 
pulsar. However the high radiation density in the vicinity of the pulsar will produce  
$e^+e^-$ pairs 
which short the 
potential drop down to values of about \E{13}. 
A different mechanisms involving Fe nuclei acceleration by relativistic MHD winds has bee put forward~\cite{Olinto3}.
But details of the effectiveness of this mechanism still need to be demonstrated.
\par{\it \bf AGN cores and jets}~:
Blast waves in AGN jets 
could in principle lead to a maximum energy of a few tens of EeV~\cite{Zas} and similarly for AGN cores.
However those maxima are unlikely 
to be achieved under realistic conditions due to the interaction of the accelerated protons with
 very high radiation fields in and around the central engine of an AGN.
\par{\it \bf FR-II radio galaxies}~:
Radio-loud quasars are characterized by a very powerful central engine ejecting matter along 
thin extended jets. 
At the ends of those jets, the so-called hot spots, the relativistic shock wave is believed to be 
able to accelerate particles up to ZeV energies. FR-II galaxies seem the best 
potential astrophysical source of UHECR~\cite{Biermann}. Unfortunately, no nearby (less than 100~Mpc) 
object of this type is visible in the direction of the observed highest energy events. 
\par{\it \bf Gamma Ray Burst}~:
Gamma ray bursters (GRB) are intense sources of gamma rays.
The most favored GRB emission model is the ``expanding fireball model''
where one assumes that a large fireball, as it expands, becomes optically thin
hence emitting a sudden burst of gamma rays. 
The observation of afterglows allowed to measure the red shift of the GRBs hence confirming their 
cosmological origin. 
GRB can be shown to accelerate protons up to \E{20}\cite{Waxman}. 
However in such a framework the UHECR spectrum should clearly show the GZK cut-off.
\subsection{Top-Down production}
\newcommand{\X}{$X$}
One way to overcome problems related to the acceleration of UHECR and the invisibility of their sources is to 
introduce a new unstable or meta-stable super-massive \X-particle. 
The decay of this \X-particle produces, among other things, quarks and leptons, 
 resulting in a large cascade of energetic photons, neutrinos and light leptons with a small 
fraction of protons and neutrons, part of which becomes the UHECR. 
For this scenario to be observable three conditions must be met:
\begin{itemize}
\item The decay must have occurred recently since the decay products must have traveled less than about 100~Mpc because of the attenuation processes discussed above.
\item The mass of this new particle must be well above the observed highest energy (100~EeV range), 
a hypothesis well satisfied by Grand Unification Theories (GUT) whose scale is around $10^{24}$-\E{25}.
\item The ratio of the volume density of this particle to its decay time must be 
compatible with the observed flux of UHECR. 
\end{itemize}
\noindent
The \X-particles may be produced by way of two distinct mechanisms:
\begin{itemize}
\item  Radiation, interaction or collapse of Topological Defects (TD), producing \X-particles that
 decay instantly. In those models the TD are leftovers
from the GUT symmetry-breaking phase transition in the very early universe.
Quantitative predictions of the TD density that survives a possible inflationary 
phase rely on a large number of theoretical hypotheses. Therefore they cannot be taken at face value,
although the experimental observation of large differences could certainly be interpreted as 
the signature of new effects.
\item  Super-massive meta-stable relic particles from some primordial quantum field, produced after
the now commonly accepted inflationary stage of our Universe. However the ratio of their lifetime 
to the age of the universe requires a fine tuning with their relative abundance. 
It is worth noting that in some of those scenarios the relic particles may also act as non-thermal 
Dark Matter.
\end{itemize}
In all conceivable Top-Down scenarios, photons and neutrinos dominate at the end of the hadronic cascade. 
This is \emph{the} important distinction from the conventional acceleration mechanisms.

\section{THE AUGER DETECTOR}
Large area ground based detectors
do not observe the incident cosmic ray directly but the Extensive Air Shower (EAS), a very large cascade of particles, 
they generate in the atmosphere. All experiments aim to measure, as accurately as possible, 
the direction of the primary cosmic ray, its energy and its nature.
There are two major techniques used. One is to build a ground array of sensors spread over a large area,
to sample the EAS particle densities on the ground.
The other 
consists in studying the longitudinal development of the EAS by detecting
the fluorescence light emitted by the Nitrogen molecules which are excited by the EAS secondaries.

The Auger Observatories\footnote{Named after the French physicist Pierre Auger} 
combine both techniques. The full design~\cite{Auger} recommends to instrument two sites,
one in the southern hemisphere (Argentina) now under contruction, and one in the north
(Utah, USA) to be started in 2003 if approved. A complete sky coverage is a mandatory condition to be
able to perform sensible analysis of the possible anisotropies of the UHECR arrival directions.
The surface of each site, 3000~km$^2$, will provide statistics of nearly 100 
events per year above 100~EeV and the detector is designed to be fully
efficient for showers above 10~EeV, with a duty-cycle of
100\%. The total aperture is 14000~km$^2$sr for both sites.

\par
Each station of the ground array is a cylindrical Cherenkov tank 
of 10~m$^2$ surface and 1.2~m height 
filled with filtered water. Three phototubes convert the Cherenkov light into electric signals 
which are digitized by means of flash ADCs running at 40MHz. Digitisation is triggered each time 
the Cherenkov ligth signal is equivalent to nearly 4 vertical muons. 
With a spacing of 1.5~km between the stations, a \E{19} vertical shower triggers on
average 6 of them, which is enough to fully reconstruct the EAS.

\par
Because of the size of the array, the stations have to work in a stand-alone
mode~: they are powered by solar panels and batteries and timing 
is provided by the GPS satellites. 
Communication with the central buildings, where the central trigger and the data collection take place,
is done through a wireless LAN.

\par
The fluorescence telescopes use photo-tubes with a field of view of $1.5^{\circ}$. 
Each telescope sees an angle of about $30\times30$
degrees. On the southern site, three eyes (7 telescopes each) will be installed
at the periphery  of the array and one (12 telescopes) in the middle, in order for
the whole array to be  visible by at least one of the telescopes. 
In hybrid mode (10\% of the events) where both the ground array and the fluorescence telescopes
 monitor the sky, the detector is expected to have on
average an energy resolution of 10\% and an angular precision better than $0.3^{\circ}$.

\begin{figure}[!t]
\includegraphics[width=14cm]{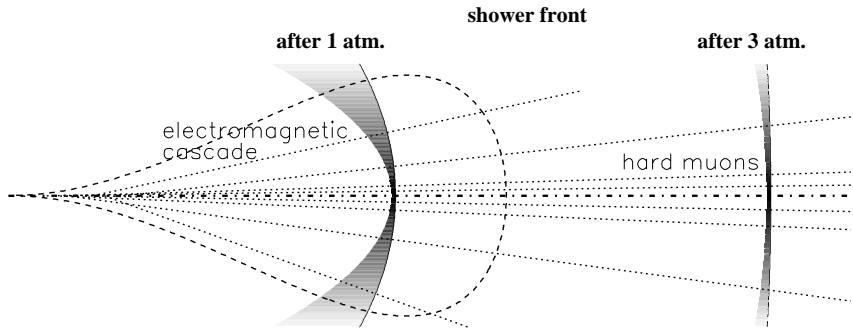}
\caption{Horizontal shower development.}
\label{showerdev}
\end{figure}

\section{NEUTRINOS}
Although both neutrinos and photons dominate the particle fluxes in Top-Down models,
only neutrinos are perfect probes of the UHECR sources. Undeflected by magnetic fields 
and with very long absorption length they allow to disantangle the characteristics for the sources from 
the propagation effects.
High energy neutrinos can also  be produced in Bottom-Up scenarios as secondaries of hadronic interaction.
If AGNs, Radio Galaxy lobes or GRBs are UHECR sources they should  produce a substantial flux 
of neutrinos. Finally, along their path in the universe, hadrons will also produce neutrinos via the pion 
photo-production processes, the so called GZK neutrinos.
\par
The first study on the detection of UHE neutrinos with the Auger detector were done by \cite{Zas2,Billoir}.
The UHE neutrinos may be detected and distinguished from ordinary hadrons by the 
shape of the horizontal EAS they produce.
At large angles, above 60$^\circ$, hadronic showers have their electromagnetic part extinguished 
(they
 have gone through a few equivalent vertical atmosphere (2 at 60$^\circ$, 3 at
70$^\circ$, 6 at 80$^\circ$)
and only high energy muons survive. 
Therefore the front of horizontal hadronic showers is very flat (the radius of curvature is larger than 100~km), 
and very narrow (the time spread is less than 50~ns). Unlike hadrons,
neutrinos may interact deeply in the atmosphere and can initiate a shower above the detector. This shower 
will appear as a ``normal''one - although horizontal -, with a curved front
(radius of curvature of a few km), a large electromagnetic component, and a wide signal 
(spread over a few microseconds) [see Figure~\ref{showerdev}].
With such important differences and if the fluxes are high enough, neutrinos can be detected and identified.
\par
Bottom-up or top-down processes hardly produce any $\nu_\tau$. In the second case one expects of course
that at the beginning of the decay chain there is a full equivalence among all flavors, but 
this symetry breaks down at the end of the fragmentation process where numerous pions are produced, 
yielding most of the expected neutrino flux. 

\par 
However, in the event of $\nu_\mu\longrightarrow\nu_\tau$ oscillations with full mixing
- a hypothesis that seems to be suported by the atmospheric neutrino data and the K2K
experiment~\cite{Kamioka-nu2000}-, one expects, for a very wide range in $\delta m^2$, to have as
many $\nu_\mu$ as $\nu_\tau$ in the cosmic ray fluxes. Unlike electrons which do not escape from 
the rocks or muons that do not produce any visible signal in the atmosphere\footnote{The electro-magnetic 
halo that surrounds very high energy muons does not spread enough in space to produce a detectable signal 
in an array of detectors separated by 1.5 km.} the taus, produced in the mountains or 
in the ground around the Auger array can escape even from deep inside the rock and produce a clear
signal if they decay above the detector.
\par
\begin{figure}[!tb]
\begin{center}
\vspace*{-1cm}
\includegraphics[width=10cm]{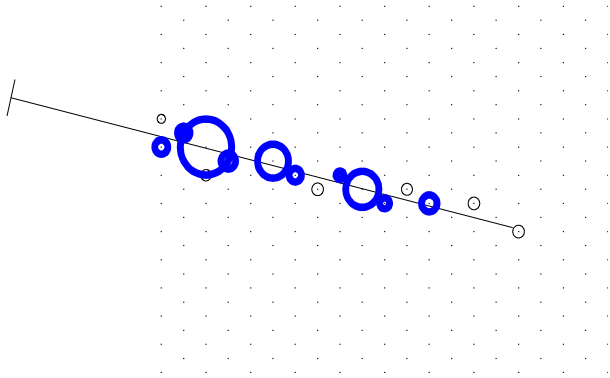}
\end{center}
\vspace*{-1cm}
\caption{Simulation of the ground trace left by a $\tau$ decay shower as produced by a 5x\E{17} tau neutrino. 
Each thick circle represents a triggered station with a surface proportional to the Cherenkov signal. 
The $\tau$ shower had an energy of 3.63x\E{17} and decayed 390 meters above the ground.}\label{tau-events}
\end{figure}

The geometrical configuration that must be met to produce a visible signal is rather severe. Neutrinos must be almost perfectly
horizontal (within about $2^\circ$) and only a few percent of the solid angle is available.
They should also interact within the decay range of the tau which must in turn have a reasonable chance of decaying 
over the size of the detector.
The neutrino energy should therefore lie within about $2\times$\E{17} 
(large flux, $\gamma c \tau \sim 10$km) and $6\times$\E{18} 
(larger average signal, $\gamma c \tau \sim 300$km). For downward going $\nu_\tau$ most of the observable 
signals come from interactions taking place
in the Andes about 50km west of the array, for upward going $\nu_\tau$ the interactions occur in the ground all around the array.
Our preliminay simulations\cite{Bertou-01} show that the maximum of the acceptance is reached for neutrino energies 
of a few \E{17} and that 80\% of the detected neutrinos are upgoing.  Figure~\ref{tau-events} shows a simulation 
of the ground trace of a tau, produced by a $3\times$\E{17} 
neutrino decaying above the ground, as sampled by the Auger stations. 
The signal is clearly visible and 10 stations pass the 4 vertical muon equivalent trigger requirement (thick circles). 
\par
\begin{figure}[!tb]
\begin{center}
\includegraphics[width=15.8cm]{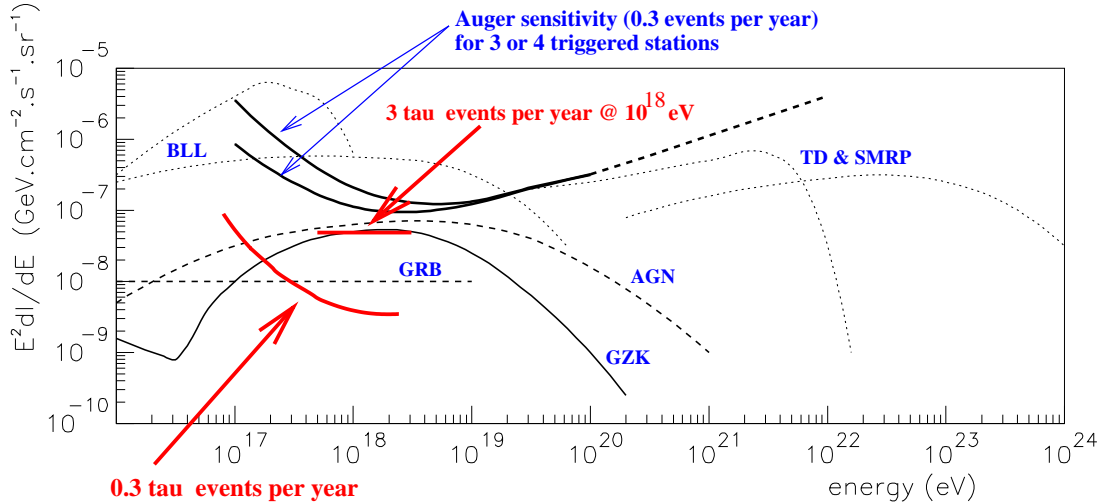}
\caption{Neutrino fluxes and ranking from various sources, dotted lines 
speculative, dashed probable, solid certain.
The two top thick solid lines represent the Auger sensitivity (0.3 event per year)
for $\nu_e$ and $\nu_\mu$ interactions in the atmosphere under two local trigger conditions. 
Sensitivity to tau induced showers are indicated by the arrows, calculation have been performed only up to neutrino energies of
$3\times$\E{18}.}
\label{nu-fluxes}
\end{center}
\end{figure}

The sensitivity limit in flux of the Auger detector - defined as the neutrino flux giving at least 1 
observed event every three years - is shown on 
Figure~\ref{nu-fluxes} together with the expected fluxes from a model calculation by Protheroe~\cite{Protheroe}.   
Both limits from atmospheric neutrino interaction~\cite{Billoir} and from $\nu_\tau$ interactions in 
the rocks~\cite{Bertou-01} are shown.
For the sensitivity to tau induced showers (as indicated by the arrows), 
the tau flux was assumed to be half of the various neutrino fluxes (full mixing hypothesis) presented on the plot. 
Only the strongest trigger requirement (4 vertical muon equivalent) was simulated. 

\par 
For standard neutrino interactions in the atmosphere, each site of the Auger observatory reaches 
10 km$^3$ water equivalent (w.e.) of target mass, and
only the models classified as speculative by Protheroe are expected to yield a detectable signal.
However, for tau induced shower the target mass is increased by a factor of nearly 100 and reaches 1000 km$^3$
w.e. at $3\times$\E{18} allowing for a detectable signal even for the lowest expected fluxes. Our preliminary
calculation shows that we can expect of the order of 3 to 6 events per year from the GZK neutrinos a very low but almost certain flux. 

\par
Only the Auger observatories can perform an optimal detection of these events. 
Above \E{17} the earth is not transparent to neutrinos and for tau or muon neutrinos successive charged current 
interactions will degrade the energy 
below \E{16} mixing the high energy signal with the more standard and far more numerous low energy one. 
Therefore a maximum of a few 100 km of rocks should intersect the neutrino trajectory to limit the number 
of interactions and allow
a high energy lepton (above \E{17}) to escape. Only nearly horizontal neutrinos interacting in 
mountains or in the top few kilometers of the Earth will allow this. 
\par
At $5\times$\E{18} and above, the tau decay length goes beyond 250 km and the probability of a 
decay over 50 km becomes small. Therefore, only the energy range $10^{17}-5\times 10^{18}$eV is available 
for detection in detectors of 
about 50km in size.  At these relatively ``low'' energies the fluorescence signal is rather small and cannot be seen 
from far away, therefore the acceptance of a detector like HiRes\cite{HiresTaup99} becomes too small.  
For the Auger ground array however, the situation is more favorable because the shower spatial size is still large 
(10 km in length and a couple of kilometers in diameter)
and because horizontal muons produce a larger signal (by a factor 3) when they cross the Cherenkov tanks.
\par
For very large detectors, like EUSO~\cite{EUSO}, the satellite space fluorescence detector, the energy threshold is above  
$5\times$\E{19} and the decay probability 
of the tau over 250km is below 10\%. Since the flux is 100 times smaller than at \E{18} and since the 
fiducial mass which is only atmosphere is still of 
the order of 1000~km$^3$ w.e. the detection of those events from space seems very unlikely.

\par
More work needs to be done to demonstrate that we can properly reconstruct the geometry of those very horizontal showers,
zenith angle and altitude above the detector, and therefore correctly estimate their energy. This is the subject of a 
paper curently in preparation\cite{Bertou-01}.

\section{CONCLUSIONS}
The origin of the highest energy cosmic ray is still not understood.
The composition, the shape of the energy spectrum and the distribution of arrivals directions of UHECR 
will prove to be powerful tools to distinguish between the different production scenarios.
\par
If UHECR are hadrons accelerated by Bottom-Up mechanisms, they should point 
back to their sources if magneic field bending is not too largre over 50 Mpc , and a  visible counterparts is expected.
They will have a specific distribution in the sky and a spectrum clearly 
showing the GZK cutoff.  
\par
For Top-Down mechanisms and above 100 EeV, one should observe a flux of photons 
and neutrinos as the photon absorption length increases (up to a few 100 Mpc). Below 
100~EeV the spectrum shape and composition  will depend on the characteristic distance between 
TD interactions or relic particle decays and Earth, the proton attenuation length and the photon absorption 
length.  
\par
In all cases, and in the hypothesis of full $\nu_\mu\rightarrow\nu_\tau$ oscillations, tau neutrinos should be visible at a rate 
which will clearly identify the UHECR production mechanisms, high rate for top down production and low rate for bottom up 
acceleration.
The detection of such a signal -but also its unexpected absence- will allow to fully constrain our models on the 
origin of the highest 
energy cosmic rays and shed some light on this mystery. The detection surface and geographic characteristics of the 
Auger observatories make them optimal detectors for such a discovery.

\font\nineit=cmti9
\font\ninebf=cmbx9

\end{document}